\documentclass{article}

\usepackage{arxiv}

\usepackage[utf8]{inputenc} 
\usepackage[T1]{fontenc}    
\usepackage{hyperref}       
\usepackage{url}            
\usepackage{booktabs}       
\usepackage{amsfonts}       
\usepackage{nicefrac}       
\usepackage{microtype}      
\usepackage{graphicx}
\usepackage{doi}
\usepackage{setspace}
\usepackage{changes}
\usepackage{amsmath}
\usepackage{float}
\usepackage{xcolor}
\usepackage{bm}
\title{Mass transfer impedance of microfluidic electrochemical chips: a semianalytical model}
\date{\textit{Revised on \today}}

\author{ \href{https://orcid.org/0000-0002-4614-1867}{\includegraphics[scale=0.06]{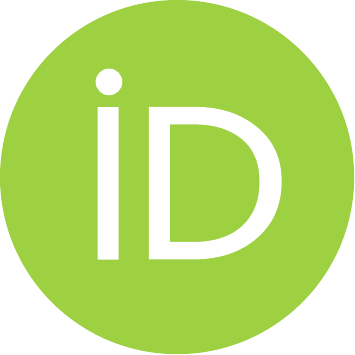}\hspace{1mm}Stéphane Chevalier}\thanks{Corresponding Author: Prof. Stéphane Chevalier, Esplanade des Arts et Métiers, 33405 Talence Cédex, FRANCE} \\
	Arts et Métiers Institute of Technology, CNRS, Université de Bordeaux, Bordeaux INP\\
	Institut de Mécanique et d'Ingénierie (I2M), Bâtiment A11,\\
	351 Cours de la Libération, 33405 Talence, France\\
	\texttt{stephane.chevalier@u-bordeaux.fr} \\
	 \AND
	 Marine Garcia \\
	Arts et Métiers Institute of Technology, CNRS, Université de Bordeaux, Bordeaux INP\\
	Institut de Mécanique et d'Ingénierie (I2M), Bâtiment A11,\\
	351 Cours de la Libération, 33405 Talence, France \\	
	 \And
	  Alain Sommier \\
	 CNRS, Arts et Métiers Institute of Technology, Université de Bordeaux, Bordeaux INP \\	 
	Institut de Mécanique et d'Ingénierie (I2M), Bâtiment A11,\\
	351 Cours de la Libération, 33405 Talence, France \\	 	 
	 \And
	 Jean-Christophe Batsale \\
	 Arts et Métiers Institute of Technology, CNRS, Université de Bordeaux, Bordeaux INP, \\
	Institut de Mécanique et d'Ingénierie (I2M), Bâtiment A11,\\
	351 Cours de la Libération, 33405 Talence, France \\	
}

\renewcommand{\shorttitle}{\textit{Submitted to Microfluid. Nanofluidics.}}

\hypersetup{
pdftitle={Impedance of a microfluidic electrochemical chip: a semianalytical approach based on integral transforms},
pdfsubject={original research paper},
pdfauthor={S. Chevalier},
pdfkeywords={Microfluidic channel, Electrochemical reaction, Mass diffusion, Current density distribution, Analytical model, Fuel cells},
}

\begin{document}
\maketitle
\clearpage
\begin{abstract}
In this paper, we report a semianalytical model for the mass transfer impedance of a microfluidic electrochemical chip (MEC). The model is based on the molar advection--diffusion equation for a microfluidic channel with a Poiseuille flow and an electrochemical reaction at the interface of deposited electrodes. Fourier--Laplace integral transforms and the quadrupole formalism are used to obtain a solution to these equations, and the three-dimensional (3D) transient concentration and current density fields are computed. This solution is validated by in-operando concentration fields measured by a visible spectroscopic imaging technique, and several equivalent electrical circuits are proposed to model mass transfer in MECs. The proposed method is the fastest way to compute the 3D transient mass transfer impedance, which can be used for a large variety of applications, such as MEC-based cytometry measurements or to predict the current density in a fuel cell.
\end{abstract}

\vspace{2cm}

\clearpage

%
%
%
%

\section{Introduction}

Microfluidic electrochemical chips (MECs) are microfabricated technologies embedded with a few microchannels through which a wide range of fluids or gazes and metallic electrodes are flowed to measure the electronic current and potential. These technologies are widely used in biology, chemistry, the food industry, medical sensors \cite{Li2021,Minami2012} and chemical production units \cite{DeMello2006}, among others. More recently, with the development of microfluidic fuel cells, MECs have been demonstrated to be good candidates for electrochemical energy converters (transforming chemical energy to electricity) \cite{Lee2012,Safdar2016}. The accuracy and performance of these technologies mainly rely on the optimization and control of the operating conditions, as well as mass transfer at the electrode interfaces. Mass transfer governs advection-diffusion-reaction processes and therefore impacts the MEC impedance \cite{Gonzalez2000}. Changes in mass transfer can significantly affect the measured current and voltage. The relationship between mass transfer and the electrode electronic conditions is the basis for most sensors for the detection of a wide range of chemicals or molecules. Thus, an in-depth understanding of mass transfer in MECs, as well as novel models and novel techniques for estimating the governing parameters, can significantly impact a large range of applications and scientific fields. \\

Modeling mass transfer in MECs is a longstanding challenge, and there is a large literature on the subject. The reader is referred to the following comprehensive reviews to find detailed studies that have been conducted on this topic over the last few years \cite{Nasharudin2014,Modestino2016,Esan2020,Honrado2021}. Several papers have focused on modeling electrode impedance (direct problem) \cite{Baltes2004,Bellagha-Chenchah2016}. Analytical models based on rules on thumb for optimizing the MEC geometry have found occasional application \cite{Perrodin2020,Moldenhauer2020}. Despite being simple, these models enable reasonably good control of MEC operating conditions. A full numerical model based on fluid dynamics and electrochemical equations has been applied in recent studies \cite{Amatore2010,Holm2019,Kjeang2007}. Numerical models capture all the physics and mechanics of the phenomena governing mass transfer, but are difficult to apply to a measurement chain or  parameter estimation. Equivalent electrical circuit (EEC) models are intermediate between models based on rules of thumb and full numerical simulations and are more suitable for on-line measurements. This approach is widely used in electrochemistry to interpret impedance spectra \cite{Arjmandi2012}. EEC models can typically reproduce experimentally observed electrochemical behavior but do not elucidate the underlying physics of the mass transfer process. There are a few studies in which EEC models have been constructed based on an advection--diffusion reaction equation \cite{Poujouly2022}, but there is still a lack of mass-transfer- based EEC for analytical or semianalytical modeling. Such tools could enable accurate modeling all three-dimensional (3D) transient mass transfer phenomena in MECs in a very short time (less than 1 s) using the EEC formalism. The lack of such models poses a considerable challenge to the development of online processing in the era of digital microfluidics \cite{Fair2007,Choi2012}, where electrodes are used ubiquitously. \\

Three-dimensional transient mass transfer in an MEC is governed by a set of partial differential equations (PDEs) that need to be solved. Among numerous available numerical methods, the fastest and most efficient for online processing is the use of analytical or semianalytical solutions. An interesting approach is the use of integral transforms, such as Laplace and Fourier transforms (which are extensively used to solve the heat diffusion equation in the textbook by Maillet et al. \cite{Maillet2000}, for example). Integral transforms have also been used in microfluidics to model mass transfer in several MEC geometries \cite{Chevalier2021b}, but a mass transfer-based EEC model has yet not been reported. Moving forward, integral transforms could be used to transform a set of PDEs into a quadrupole formalism for use in formulating an EEC model for the mass transfer impedance. Many studies based on quadrupole approaches can be found in the thermal sciences \cite{Bendada1998,Aouali2021}, and the use of these approaches has been crucial for the extensive development and improvement of inverse methods. Therefore, the similarities between PDEs for mass and heat transfer can be exploited
 to derive a quadrupole model and an associated EEC model of 3D transient mass transfer in MECs. The derivation of such a model is the goal of this study.\\

First, 3D transient equations for mass transfer in an MEC are established. The use of the integral transform is described in detail to relate the PDE equations, quadrupole formalism and EEC proposed in the paper. Given the numerous assumptions made to develop the model, an experimental setup was developed to measure the concentration fields in an operating MEC \cite{Garcia2022}. In the results and discussion section of the paper, the model is validated against experimental data, and several EECs are used to describe mass transfer. An example of a mass transfer calculation and a discussion of the model performance for an MEC with an electrode array are provided at the end of the paper. In this study, pioneering methods developed in inverse engineering science for heat transfer are applied to model microfluidic technology for use in improved sensors or chemical reactors.\\

\section{Methods}
\subsection{Electrochemical impedance of a single electrode}
The objective of this study is to develop an analytical solution for the electrochemical impedance of an electrode in an MEC. The 3D transient impedance $Z$ (m$^{-1}$.s$^{-2}$) between the Faradic current and the limiting electrode concentration at the electrode surface is defined as
\begin{equation}
j_F(x,y,t) = z_eF(Z(x,y,t)\otimes c_{lim}(x,y,t)), \label{e_def_impedance}
\end{equation}
where $j_F$ is the Faradic current density (A/m$^2$), $z_e$ is the number of electrons exchanged during the reaction, $F$ is the Faraday constant (C/mol), and $c_{lim}$ (mol/m$^3$) is the electrode limiting concentration (either 0 or $c_0$, the final product concentration). The operator $\otimes$ stands for the convolution product.\\

A convenient way to compute a convolution product is to use integral transforms, i.e., Fourier and Laplace transforms, for which the convolution product is a simple product. Thus, Equation \ref{e_def_impedance} can be rewritten as
\begin{equation}
j_F(x,y,t) = z_eFc_0\mathcal{T}^{-1}\lbrace\mathcal{Z}(p,\alpha_n,s)\gamma_{lim}\rbrace,\label{e_exp_current_faradic}
\end{equation}
where $\mathcal{T}^{-1}$ is the inverse Fourier--Laplace transform; $p$ and $s$ are the Laplace parameters of $x$ and $t$, respectively; $\alpha_n$ are the spatial frequencies in $y$; $\mathcal{Z}$ (m/s) is the transformed impedance, and $\gamma_{lim}$ (mol.s/m) is the transformed dimensionless limiting electrode concentration. Equation \ref{e_exp_current_faradic} can be used to determine the transient Faradic current distribution at the interface of an electrode in a three-dimensional MEC. \\

The methodology and the mathematical tools used to determine the function $\mathcal{Z}(p,\alpha_n,s)$ are described in the following sections. Most of the methodology used in this paper is based on the thermal quadrupole formalism classically used in heat transfer to model heat diffusion. The reader is referred to publications by Maillet et al. and Bendada et al. \cite{Maillet2000,Bendada1998} for mathematical details on quadrupole construction.

\subsection{Mass transfer modeling}
In this study, impedance modeling is carried out using a commonly used electrode geometry, i.e., plane rectangular electrodes deposited at the bottom of the channel \cite{Ibrahim2022}. The impedance of a single electrode is derived before generalizing the model to any electrode array.

\subsubsection{General equations}

\begin{figure}[H]
\centering
\includegraphics[scale=.5]{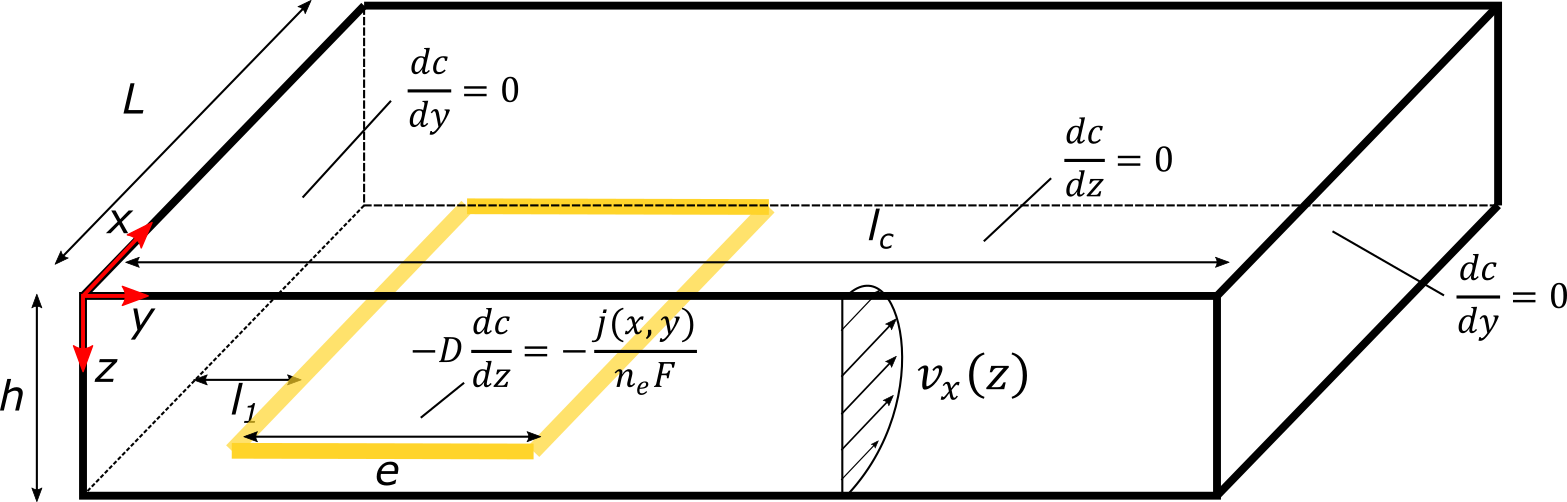}
\caption{A 3D schematic of the channel with a single electrode placed at $z=0$ (in yellow). The boundary conditions are also indicated.}
\label{f_schema_3D}
\end{figure}

The 3D geometry of the MEC with a single electrode is presented in Figure \ref{f_schema_3D}. In this model, the velocity profile is considered laminar and fully developed in the $x$-direction, i.e., Poiseuille flow. The mass diffusivity is considered constant, which is justified for dilute fuels in aqueous solutions, and Fick's law is used to model mass transport. Charge transfer is modeled using the Tafel law at the electrode interface, and ohmic losses are negligible because the electrodes produce a very low current density (a few µAs).\\

Under these conditions, the 3D problem of mass transport can be written as

\begin{equation}
\frac{\partial c}{\partial t}+ v_x(y,z)\frac{\partial c}{\partial x}= D\left(\frac{\partial^2 c}{\partial x^2}+\frac{\partial^2 c}{\partial z^2}+\frac{\partial^2 c}{\partial y^2}\right), \label{e_transport_masse_3D}
\end{equation}
where $c$ is the reactant molar concentration (M); $D$ is the mass diffusivity (m$^2$/s); $x$, $y$, and $z$ are the spatial coordinates (m); and $t$ is the time (s). The current density produced or consumed at the electrode interface is modeled using the Tafel law, which relates the local reactant concentration at the electrode interface to the fuel cell potential as
\begin{equation}
j(x,y,t) =i_0\frac{c(x,y,z=0,t)}{c_0}\exp(\eta(t)/b), \label{e_tafel}
\end{equation}
where $j(x,y,t)$ is the current density distribution (A/m$^2$) on the electrode surface (the current density is null outside the electrode), $i_0$ is the electrode exchange current (A/m$^2$), $b$ is the Tafel slope (V) and $\eta$ is the overpotential (V) that depends on the electrode potential. Finally, the velocity profile, $v_x(y,z)$, can be written analytically assuming a Poiseuille velocity profile in a rectangular channel as \cite{Chevalier2021,Bruus2008}
\begin{equation}
v_x(y,z) = \frac{4h^2\Delta p}{\pi^3\mu L}\sum_{k,odd}^\infty\frac{1}{k^3}\left[1-\frac{\cosh(k\pi\frac{2y-l_c}{2h})}{\cosh(k\pi\frac{l_c}{2h})}\right]\sin\left(k\pi\frac{z}{h}\right),
\label{e_v_profile}
\end{equation}
where $h$, $L$ and $l_c$ are the channel dimensions indicated in Figure \ref{f_schema_3D}, $\Delta p$ is the pressure difference (Pa), $\mu$ is the viscosity (Pa.s) and $k$ an integer.

\subsubsection{Solution using integral transforms}

Several assumptions are made to formulate a simplified model containing only the most important phenomena and obtain an analytical solution. The aspect ratio of the channel, i.e., $\varepsilon=l_c/h$ is considered to be sufficiently large (higher than 10) that the velocity profile in the $y$-direction can be considered constant \cite{Chevalier2021}, such as $v_x(z)=6v_{moy}\left(z-z^2/h\right)/h$. The Peclet number in the $x$-direction is considered to be sufficiently large that diffusion in this direction can be neglected, i.e., $Pe\gg 1$ and $\partial^2c/\partial x^2\approx 0$. The chemical concentration is assumed to be constant at the channel entrance (there is no mixing with another fluid) and equal to $c(x=0,y,z,t)=c_0$ in this study. The chemical species is considered to be consumed along the electrode length, and the boundary condition at the electrode interface is modeled using Faraday's law as $\left.\partial c/\partial z\right|_{z=h}=j(x,y,t)/(z_eF), \label{e_CL_courant}$, where $F$ is the Faraday constant and $z_e$ is the number of electrons exchanged during the reaction.\\



To solve for the 3D transient concentration field, we define the following integral transform:
\begin{equation}
\gamma(p,\alpha_n,z,s)=\int_0^{+\infty}\int_0^{+\infty}\int_0^{l_c}\Delta \tilde{c}(x,y,z,t)\cos(\alpha_n y)e^{-px-st}dydxdt,\label{e_int_transf}
\end{equation}
where $\gamma$ (mol.s/m) is the transformed dimensionless concentration, with $\alpha_n=n\pi/l_c$, $n\in[0,N]$ and $N$ being the number of spatial frequencies (ideally infinite). The dimensionless concentration $\Delta\tilde{c}=(c-c_0)/c_0$ is used to ensure a zero initial condition for the Laplace transforms. Thus, substituting Equation \ref{e_int_transf} into Equation \ref{e_transport_masse_3D} and using the aforementioned assumptions yields a transformed system of equations as
\begin{eqnarray}
\frac{d^2\gamma}{d z^2}-\kappa_i^2\gamma&=&0,\label{e_ODE_trans}\\
\left.\frac{d \gamma}{d z}\right|_{z=0}&=&0,\\
\left.-D\frac{d \gamma}{\partial z}\right|_{z=h}&=&-k_0(1-e^{-Lp})\int_0^l f(y)\left(\Delta \bar{c}(p,y,z=h,s)-\frac{1}{ps}\right)\cos(\alpha_n y)dy,
\label{e_convolution_conv_conc}
\end{eqnarray}
where $k_0$ is the reaction rate constant defined as $k_0 = i_0e^{\eta/b}/(z_eFc_0)$ (see Equation \ref{e_tafel} and Faraday's law), $L$ is the length of the electrode, and $\kappa_i=\sqrt{\alpha_n^2+v_ip/D+s/D}$ (m). The function $f(y)$ describes the spatial geometry of the electrode in the $y$-direction.\\

The major difficulty in the last equation is the computation of the boundary condition \ref{e_convolution_conv_conc}. This boundary condition corresponds to the convolution product between the spectrum of the concentration at the electrode interface, $\gamma(h)$, and the spectrum of the electrode geometry in the $y$-direction, $f(y)$. Computation of this convolution operation is detailed in the appendix.\\
Finally, the last set of equations is a trivial ordinary second-order differential system, for which there is  a quadrupole representation in terms of matrix representation or impedances \cite{Maillet2000}. This system can be related to the concentration and molar flux at $z=0$ and $z=h$ by using a transfer matrix, such as




\begin{equation}
\begin{pmatrix}
 \bm{\gamma}(0)\\
\bm{\dot{N}}(0)=0
\end{pmatrix}
=
\begin{bmatrix}
\bm{\mathcal{A}}&\bm{\mathcal{B}}\\
\bm{\mathcal{C}}&\bm{\mathcal{D}}
\end{bmatrix}
\begin{bmatrix}
\bm{I}&\bm{K}^{-1}\\
\bm{0}&\bm{I}
\end{bmatrix}
\begin{pmatrix}
\bm{\gamma}_{lim}\\
\bm{\dot{N}}(h)
\end{pmatrix},
\label{e_quadripole_vit_discret}
\end{equation}

where $\bm{\dot{N}}(0)$ and $\bm{\dot{N}}(h)$ are the transformed molar flux at $z=0$ and $z=h$, respectively, and $\bm{\gamma}_{lim}$ is the transformed limiting concentration at the electrode (see Section 2.1). Note that $\bm{\dot{N}}(0)$ and $\bm{\dot{N}}(h)$ represent the boundary conditions $-D\left.\frac{d \gamma}{d z}\right|_{z=0}$ and $\left.-D\frac{d \gamma}{\partial z}\right|_{z=h}$, respectively. All the matrices $\bm{\mathcal{A}}$, $\bm{\mathcal{B}}$, $\bm{\mathcal{C}}$, and $\bm{\mathcal{D}}$ are diagonal, and the matrix $\bm{K}$ represents the boundary condition \ref{e_convolution_conv_conc}. The matrix $\bm{I}$ is an identity matrix of size $N$.\\
Using the quadrupole formalism offers the considerable advantage of enabling serial aggregation in the $z$-direction of an infinity of domains with different transport coefficients. These properties can be used to take into account the change in the velocity in the $z$-direction. In the present case, the microchannel is discretized into $M$ small channels with different velocities, $v_i$. This property is used to compute the $\bm{\mathcal{A}}$, $\bm{\mathcal{B}}$, $\bm{\mathcal{C}}$, and $\bm{\mathcal{D}}$ matrices as
\begin{equation}
\begin{bmatrix}
\bm{\mathcal{A}}&\bm{\mathcal{B}}\\
\bm{\mathcal{C}}&\bm{\mathcal{D}}
\end{bmatrix}
=
\prod_{i=1}^M
\begin{bmatrix}
\bm{A}_i&\bm{B}_i\\
\bm{C}_i&\bm{A}_i
\end{bmatrix},\label{e_coef_mat_velocity}
\end{equation}

where $\prod$ is the product operator. The coefficients of the quadrupole matrix for each small channel are defined as
\begin{eqnarray}
\bm{A}_i&=&\text{diag}(\cosh(\kappa_i dh))\label{e_coef_Ai}\\
\bm{B}_i&=&\text{diag}(\sinh(\kappa_i dh)/D/\kappa_i)\\
\bm{C}_i&=&\text{diag}(D\kappa_i\sinh(\kappa_i dh))\label{e_coef_Ci}\\
\kappa_i&=&\sqrt{\alpha_n^2+pv_i/D+s/D},
\end{eqnarray}

where $dh=h/(M-1)$ is the thickness of the discrete small channel elements, and $v_i$ is the velocity in the small channel element calculated from Equation \ref{e_v_profile}, i.e., $v_i=v((i-\frac{1}{2})dh)$ with $i=1$ to $M$. Note that given the symmetry of the velocity profile in $z$, it can be shown that $\bm{\mathcal{D}}=\bm{\mathcal{A}}$.\\
Thus, Equation \ref{e_quadripole_vit_discret} can be used to obtain all the distributions of interest, such as the electrochemical impedance, the molar flux at the electrode interface, and the local molar concentration in the channel:
\begin{eqnarray}
\bm{\mathcal{Z}} &=& [\bm{\mathcal{A}}\bm{\mathcal{C}}^{-1}+\bm{K}^{-1}]^{-1}, \label{e_impedance_local}\\
\bm{\dot{N}}(h) &=& \bm{\mathcal{Z}}\bm{\gamma}_{lim},\\
\bm{\gamma}(z) &=& \bm{\mathcal{A}}(z)\bm{\mathcal{A}}^{-1}(\bm{I}-\bm{K}^{-1}\bm{\mathcal{Z}})\bm{\gamma}_{lim}.\label{e_conc_local}.
\end{eqnarray}
In Equation \ref{e_conc_local}, $\bm{\mathcal{A}}(z)$ is the matrix obtained from Equation \ref{e_coef_mat_velocity} for $M=z/dh+1$. Equation \ref{e_conc_local} can be simplified using the diagonality of the matrices as $\bm{\mathcal{A}}-\bm{\mathcal{A}}\bm{K}^{-1}\bm{\mathcal{Z}}-\bm{\mathcal{B}}\bm{\mathcal{Z}}=\bm{\mathcal{A}}[\bm{I}-\bm{K}^{-1}\bm{\mathcal{Z}}]-\bm{\mathcal{B}}\bm{\mathcal{Z}}\approx \bm{\mathcal{A}}^{-1}[\bm{I}-\bm{K}^{-1}\bm{\mathcal{Z}}]$. This property can be generalized to any matrix at a position $z$.\\

Finally, the transformed distributions can be computed in the $x,y,z,t$ space using the following expression:
\begin{equation}
\mathcal{O}(x,y,z,t) = \mathcal{L}^{-1}_{x,t}\left\lbrace\frac{1}{l_c}\mathcal{O}_0(p,\alpha_n,z,s)+\frac{2}{l_c}\sum_{n=1}^N\mathcal{O}_n(p,\alpha_n,z,s)\cos(\alpha_n y)\right\rbrace,
\end{equation}
where $\mathcal{O}$ denotes one of the distributions mentioned above, the indices $n$ are the spatial frequencies, and $\mathcal{L}^{-1}_{x,t}$ is the double inverse Laplace transform in $x$ and $t$. The inverse transforms are computed numerically using either the algorithm proposed by Stehfest or that proposed by Den Iseger \cite{Stehfest,DenIseger2006}.\\
Finally, another considerable advantage of the quadrupole formalism can be seen in the straightforward calculation of the total Faradic current produced or consumed in an MEC:

\begin{equation}
I_F(t) = \int_0^L\int_0^{l_c}j_F(x,y,t)dydx=c_0z_eF\int_0^L\mathcal{L}_{x,t}^{-1}\lbrace\dot{N}_0(h)\rbrace dx, \label{e_courant_tot}
\end{equation}
where $\dot{N}_0(h)$ is the first component of the molar flux spectrum.


\subsection{Experiments}
A T-shaped microfluidic channel was fabricated using standard photolithography. A negative photoresist was spin coated on a silicon wafer, covered with a photomask and exposed to UV light. The photoresist was then submerged in a propylene glycol methyl ether acetate (PGMEA) solution for development. The obtained mold was placed in a Petri dish and coated with 5 mm of polydimethylsiloxane (PDMS). The PDMS was cured, peeled off the mold and hole-punched to create two inlets and one outlet. The microchannel had a height of 25 µm, a width of 3 mm, and a length of 15 mm. This specific aspect ratio was used to simplify the MFC model (see the following section).\\

\begin{figure}[H]
\centering
\includegraphics[scale=.6]{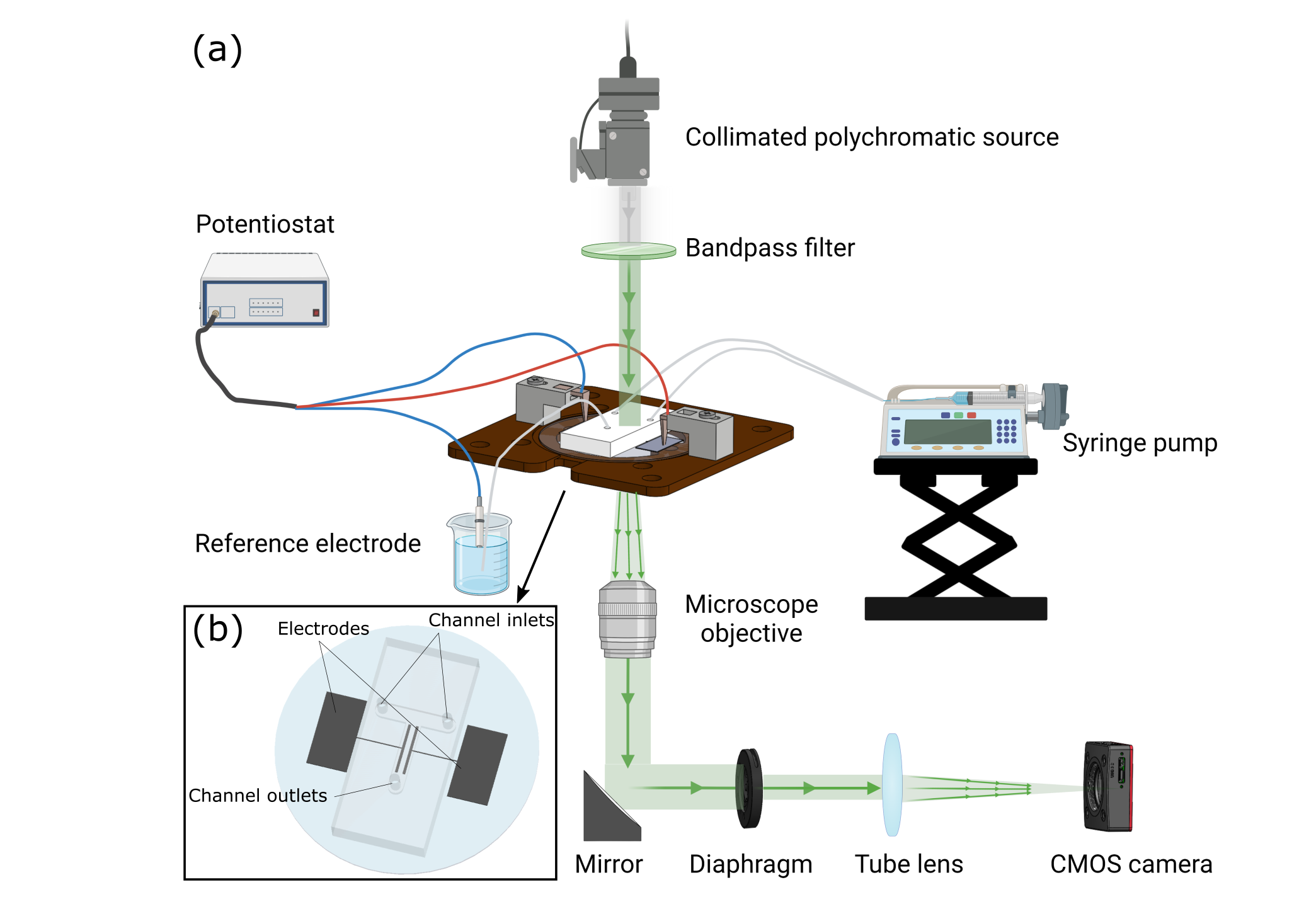}
\caption{(a) Schematic of the visible spectroscopic imaging setup used to measure the in operando concentration. (b) 3D view of the MFC used to validate the model.}
\label{f_setup}
\end{figure}

To fabricate the electrodes, an inverse pattern was first created on a glass wafer using the same photolithography process employed for the PDMS stamp. The electrodes were deposited by sputtering ~60 nm of titanium as an adhesion layer followed by ~300 nm of platinum as the catalyst material. The remaining photoresist was removed by submerging the wafer in a chemical etching solution to obtain only the platinum pattern directly on the glass substrate. The electrode had a width of 500 µm and a length of 10 mm. The PDMS stamp was plasma activated and bound to the glass substrate, yielding the complete MFC presented in Figure \ref{f_setup}.\\

The concentration distribution and the total current produced by the MFC were measured using the setup shown in Figure \ref{f_setup}. The setup is a homemade inverse microscope. White collimated light is used as the primary light source. A narrow bandpass filter ($\lambda=540\pm 5 $ nm) is used to produce a monochromatic green light passing through the MFC. The light is finally collected through a microscope objective and a lens to produce an image on a CMOS camera (Zelux 1.6 MP Color CMOS Camera). Only the green channel of the camera is used for image postprocessing.\\

The MFC was controlled using a potentiostat to measure the voltage and the current produced. An Ag/AgCl reference electrode was used to measure both anode and cathode potentials. The reactant flow rate was controlled using a syringe pump at a constant value of 0.5 µl/min.\\

The reactants (formic acid and potassium permanganate) were chosen to ensure good MEC performance \cite{Kjeang2009}. In addition, potassium permanganate has a clear absorption signature in the visible range and therefore offers a considerable advantage over formic acid, which is transparent. The oxidation reaction of formic acid at the anode is
\begin{equation}
HCOOH \longrightarrow CO_2+2H^++2e^-,
\end{equation}
and the reduction reaction for permanganate at the cathode is
\begin{equation}
MnO_4^-+8H^+ +5e^- \longrightarrow Mn^{2+}+4H_2O.
\label{e_manganese}
\end{equation}
Equation \ref{e_manganese} shows current production transforms the permanganate ions ($MnO_4^-$) into manganese ions $Mn^{2+}$. Thus, the application of a current through the MFC electrode triggers a decrease in the permanganate concentration, which can be measured by visible spectroscopy.\\

The wavelength chosen in the imaging setup corresponds to the strongest light absorption of the permanganate ions, because formic acid is completely transparent. The Beer-Lambert equation can be used to relate the variation in light intensity to the variation in the permanganate concentration as
\begin{equation}
\Delta c = -\alpha^{-1}\log_{10}\left(\frac{I_0+\Delta I}{I_0}\right),
\label{e_BL}
\end{equation}
where $\alpha$ is the permanganate absorption coefficient (mM$^{-1}$), $I_0$ is the light intensity of the background and $\Delta I$ is the variation in the light intensity induced by the decrease in  the permanganate concentration. The permanganate absorption coefficient was measured to be $\alpha = 5,5\times 10^{-3}$ mM$^{-1}$ at 540 nm for a channel thickness of 25 µm.

\section{Results and discussion}

\subsection{Validation of the model in steady state}

\begin{figure}[H]
\centering
\includegraphics[scale=.3]{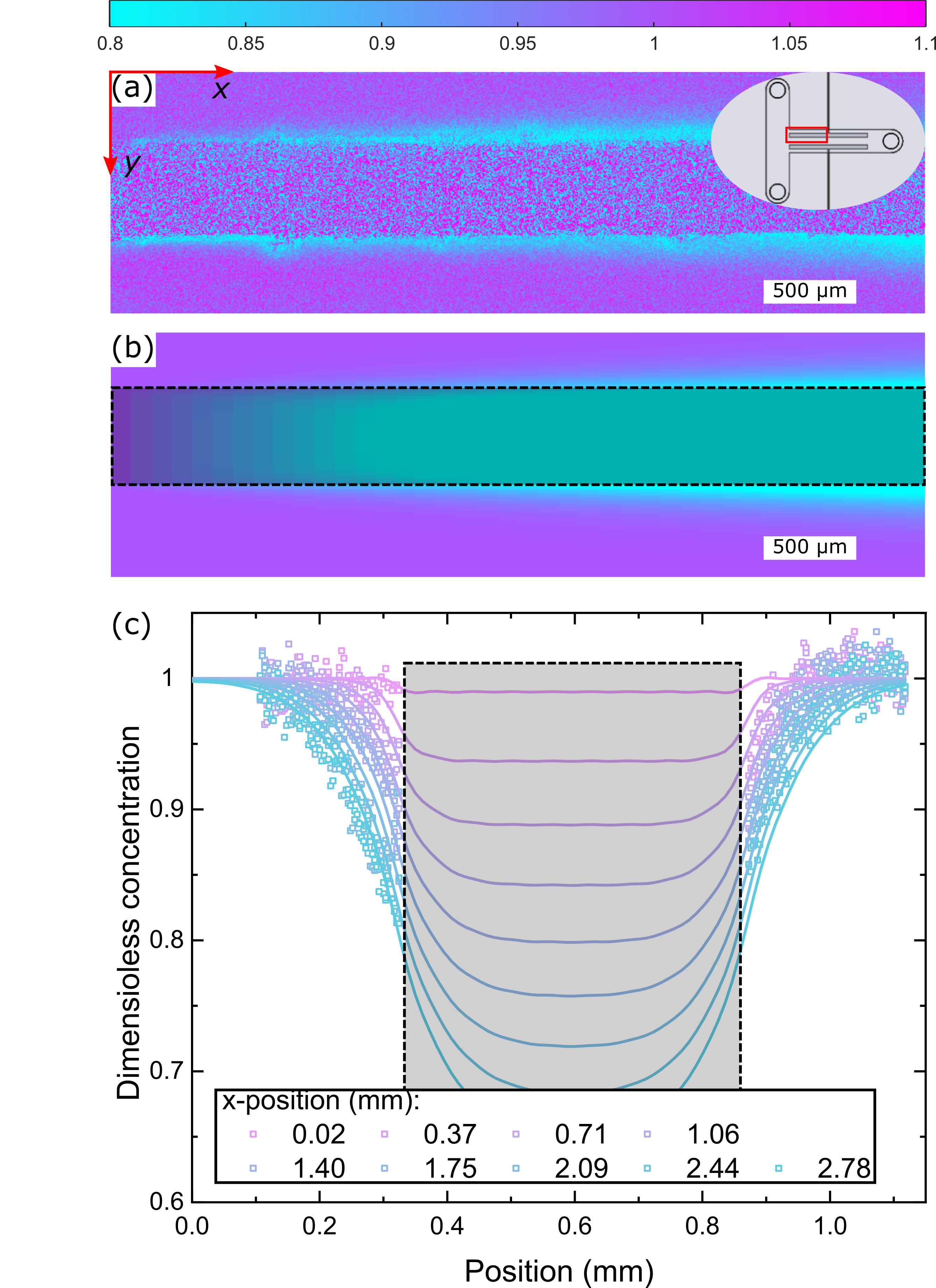}
\caption{Average dimensionless concentration distribution, $c(x,y)/c_0$. (a) The measured concentration at steady state. The inset shows the region in the cell where the image was taken (red rectangle). (b) The modeled concentration obtained after the fitting process. The dashed rectangle represents the electrode area. (c) Comparison of the measured and modeled profiles for a range of $x$-positions.}
\label{f_comp_model}
\end{figure}

 The model was validated by imaging the concentration field in an operating MEC. A current of 20 µA was produced, resulting in a decrease in the permanganate concentration at the electrode interface (as shown by Equation \ref{e_manganese}). Once the steady state was reached, i.e., after 30 s, the concentration field was imaged for 2 s and averaged to increase the signal-to-noise ratio. More details on the image processing, noise analysis and visible spectroscopic imaging technique can be found in \cite{Garcia2022}. The experimental data are presented in Figure \ref{f_comp_model}(a). The permanganate concentration was initially 10 mM at the channel inlet (shown on the left in Figure \ref{f_comp_model}(a)) and decreased along the $x$-direction on both sides of the electrode. The diffusion of the concentration depletion in the $y$-direction can also be seen in the figure. There was no signal in the electrode area because the metal completely absorbed visible light.\\

To validate the model, the experimental data is fit to determine two unknown parameters, i.e., the mass diffusivity $D$ and the reaction rate constant $k_0$. All the other geometrical data and operating conditions used in the experiment were known and set in the model. The simplex algorithm from MATLAB was used to estimate $D$ and $k_0$ by minimizing the difference between the computed concentration and the concentration measured experimentally on both sides of the electrodes. The calculation of the concentration field $c(x,y,z=h)$ took approximately 1 s for several $x$ and $y$ positions on a notebook laptop (Intel Core i7-8550U CPU at 1.80 GHz and 16 GB of RAM). Therefore, the estimation procedure for the parameter values took less than 1 min to converge. This result demonstrates how rapidly a minimization algorithm can be executed by using semianalytical mass transfer models. Analytical approaches also offer a considerable advantage over numerical solution in that the concentration, molar flux or current density can be computed at a single location without solving the problem in the entire space and time domain (which can be quite time-consuming for a 3D transient problem).\\

Execution of the minimization algorithm yielded the fitted parameters as $D=(1.24\pm0.10)\times 10^{-3}$ mm$^2$/s and $k=(1.30\pm0.04)\times 10^{-3}$ mm/s. Using this parameter set resulted in a total current density obtained from Equation \ref{e_courant_tot} of 19.8 µA, which is in very good agreement with the experimentally measured current, i.e., 20 µA. The resulting concentration field computed using Equation \ref{e_conc_local} at $z=h$ is shown in Figure \ref{f_comp_model}(b). Qualitatively good agreement is observed between the theoretical result and experimental data, that is, the same trends in the diffusion and concentration consumption are observed. The more detailed comparison presented in Figure \ref{f_comp_model}(c) shows that exactly the same concentration profile was obtained by the model as from the experiments. This result validates the analytical development and assumptions made in the present study. This result also shows how using the present analytical model in conjunction with the in operando concentration field measurements in an MEC provides a powerful tool for characterizing mass transfer. Under these conditions, the Damköhler number was found to be $Da=k_0l_c^2/(hD)\approx 380$. The Peclet number was $Pe\approx 800$, which validates the assumption of negligible diffusion in the $x$-direction.

\subsection{Impact of the operating conditions and MEC design on mass transfer}

In this section, the effect of two important parameters on mass transfer is investigated: the channel aspect ratio and inlet flow rate. The objective is to determine the regimes in which the impedance model proposed in Equation \ref{e_impedance_local} can be simplified.\\

\begin{figure}[H]
\centering
\includegraphics[scale=.3]{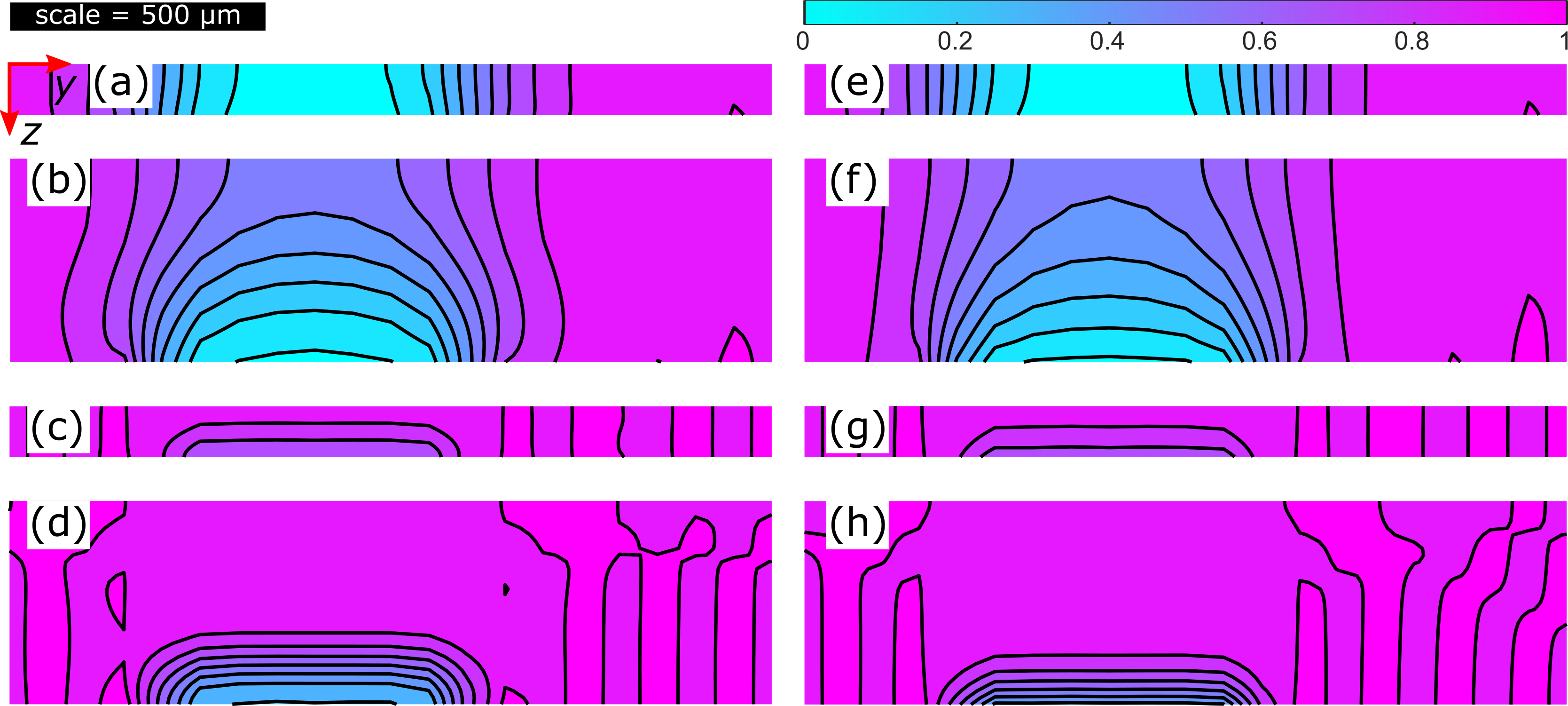}
\caption{Dimensionless steady-state concentration distribution at $L/2$, i.e., $c(x=L/2,y,z)$, between $y=0$ and $y=l_c/2$ computed using the same 3D model and operating conditions with (a)-(d) the local velocity profile (at two Peclet numbers, Pe = 800 (a)-(b) and Pe = $2\times10^4$ (c)-(d)) and (e)-(f) with a constant velocity. The channel aspect ratio varies from 15 to 60 (the aspect ratio in the figure was multiplied by 2 in $z$ for clarity).}
\label{f_diag_regime}
\end{figure}

Figure \ref{f_diag_regime}(a) to (d) shows the concentration fields in the $y$- and $z$-directions calculated using Equation \ref{e_conc_local}. These concentrations are shown at $x=L/2$, i.e., the middle of the channel, and at steady state. A total of $N=41$ spatial frequencies were used, and the channel was discretized into 31 layers ($M=31$). It took approximately 3 s to compute the 2D field on the same laptop. The Damköhler number was maintained constant at 2000 in all simulations. Only the channel aspect ratio (through the value of $h$) and the inlet flow rate, $q_{tot}$, were varied. Figure \ref{f_diag_regime}(a) for the case of the highest aspect ratio and the lowest flow rate shows that the MEC can be considered to be 2D because the concentration profile does not depend on $z$. This scenario corresponds to the experimental case presented in the previous section and justifies computing only $\gamma(h)$ to accurately estimate the mean concentration in the channel.\\

For all the other cases presented in Figures \ref{f_diag_regime}(b) to (d), the 3D aspect of the channel needs to be taken into account by discretization in the $z$-direction (see Equation \ref{e_coef_mat_velocity}). Even for a high channel aspect ratio, a large flow rate results in a significant gradient in the $z$-direction. Another difference between these three concentration fields stems from the use of a parabolic velocity profile $v(z)$. For comparison, the concentration profiles shown in Figures \ref{f_diag_regime}(e) to (g) were obtained by using the same simulation procedure with a constant velocity profile, $v(z) = v_{moy}$. Differences between Figures \ref{f_diag_regime}(d) and (h) are only noticeable at a low channel aspect ratio and a high inlet flow rate (a Peclet number larger than 10$^3$). Under these conditions, the velocity boundary layer considerably impacts the diffusion process.\\

The effect of the parabolic velocity profile on the concentration field can be elucidated by comparing two boundary layers: diffusive and laminar. The size of a diffusive boundary layer can be approximated by $\tilde{\delta}_D\sim Pe^{-1/2}$, whereas that of a laminar boundary layer can be approximated to first order by Lévêque theory as $\tilde{\delta}_{Lev}\sim(3\pi\varepsilon)^{-1}$ \cite{Chevalier2021b}. Inside the laminar boundary layer, i.e., close to the electrode interface, the diffusion gradient is hindered, leading to reduced molar rates \cite{Braff2013} with a scaling law of $\sim Pe^{-1/3}$. However, this effect vanishes once the diffusion boundary layer becomes larger than $\tilde{\delta}_{Lev}$. Therefore, the definitions of these boundary layers show that the velocity profile has the strongest effect on the concentration field for a low channel aspect ratio and a high Peclet number, leading to $\tilde{\delta}_{Lev}>\tilde{\delta}_D$. This case is shown in Figure \ref{f_diag_regime}(d).\\

However, particular attention must be paid to the case of a small aspect ratio for which the laminar boundary layer in the y-direction grows significantly. In this case, the assumption of a parabolic velocity profile may not be valid close to the lateral walls, i.e., $y\sim 0$(see, for example, ref \cite{Chevalier2021b}), and the average velocity can differ from $v_{moy}$. Therefore, the value of $v_{moy}$ needs to be analyzed and validated for $\varepsilon<10$.\\

\subsection{Using an equivalent electrical circuit to model mass transfer}

\begin{figure}[H]
\centering
\includegraphics[scale=.27]{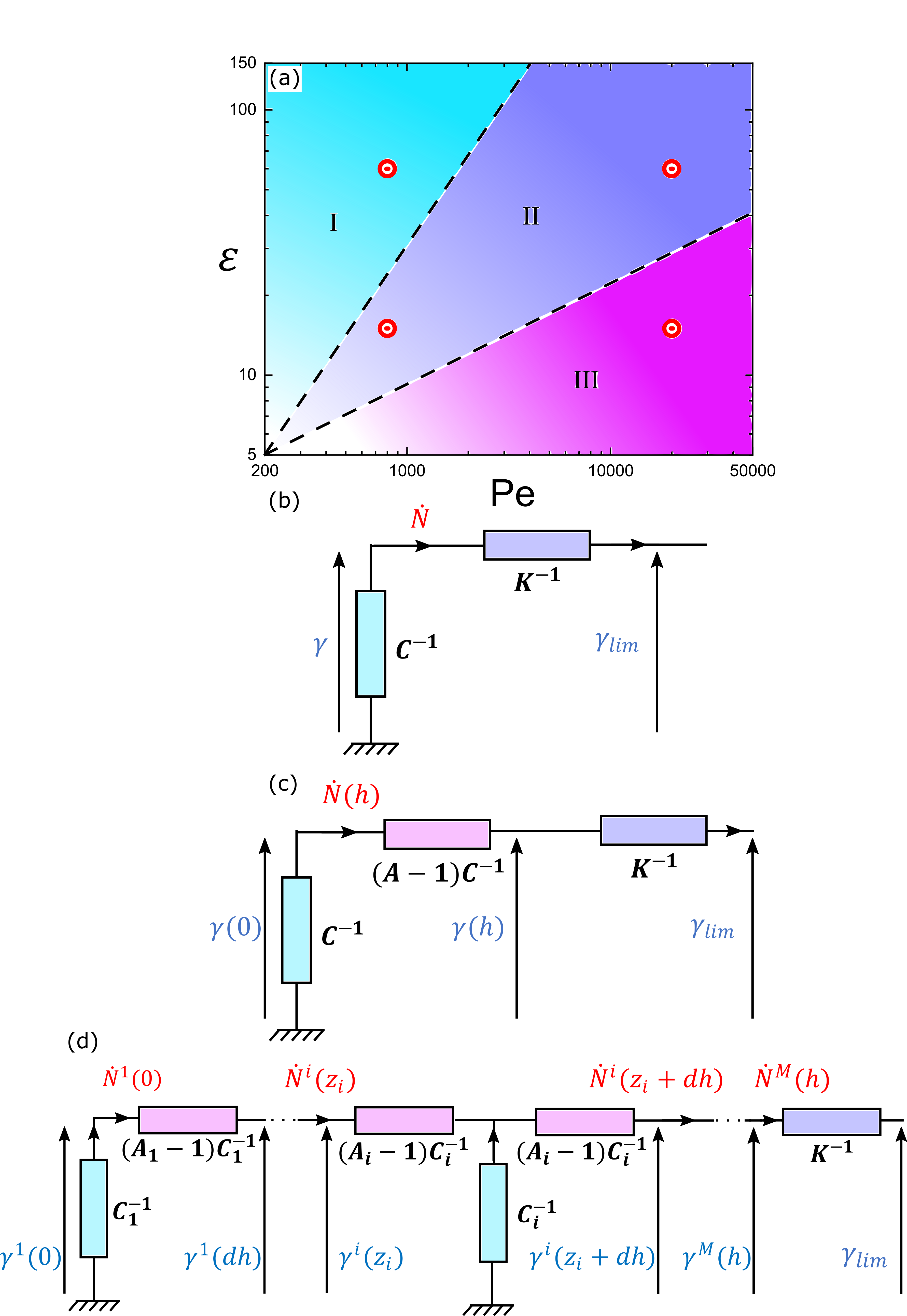}
\caption{(a) The phase diagram of the different mass transfer regimes in the $\varepsilon$ - Pe plane. (b), (c) and (d) EECs corresponding to Regimes I, II and III, respectively. The matrices $\bm{A}$ and $\bm{C}$ are computed using the definitions of $A_i$ and $C_i$ with a constant velocity, i.e., $v_i=v_{moy}$.}
\label{f_impedance}
\end{figure}

In the results presented in Figure \ref{f_diag_regime}, several mass transfer regimes can be identified depending on the aspect ratio and Peclet number. Three regimes are indicated in Figure \ref{f_impedance}(a). The red symbols correspond to the operating conditions used to compute the results presented in Figure \ref{f_diag_regime}. The dashed boundaries between the three regimes were then generated  based on these results
.\\

In Regime I, the channel aspect ratio is high and the Peclet number is low, and mass transfer can be computed in 2D, i.e., as there is a negligible concentration gradient in the $z$-direction, $\gamma(0)\approx\gamma(h)$ and the mean velocity $v_{moy}$ can be used instead of the parabolic profile $v(z)$. In this case, the EEC model presented in Figure \ref{f_diag_regime}(b) can be used. The impedance is thus simplified to $\bm{\mathcal{Z}_{2D}}=(\bm{C}^{-1}+\bm{K}^{-1})^{-1}$. The average concentration and molar rates are then deduced as $\gamma_{2D}=(\bm{I}+\bm{K}^{-1}\bm{C})^{-1}\bm{\gamma}_{lim}$ and
$\bm{\dot{N}}_{2D}=\bm{\mathcal{Z}_{2D}}\bm{\gamma}_{lim}$, respectively. The matrix $\bm{C}$ is computed using Equation \ref{e_coef_Ci} in the case of a constant velocity. Regime I corresponds to the experimental conditions presented in Section 3.1.\\

In Regime II, as both the channel aspect ratio and Peclet number are high, the mass transfer needs to be computed in 3D, but the mean velocity can still be used. The EEC model in this regime is presented in Figure \ref{f_diag_regime}(c). In this case, $\gamma(0)\neq\gamma(h)$, and the matrices in Equations \ref{e_coef_Ci} and \ref{e_coef_Ai} are computed using a constant velocity. The impedance, molar concentration and molar flux are then obtained using Equations \ref{e_impedance_local} to \ref{e_conc_local}.\\

Regime III corresponds to the general case where the three-dimensionality of the channel and the parabolic velocity profile need to be taken into account. Such a case arises for a small channel aspect ratio and a large Peclet number. The equivalent electrical circuit of the mass transfer impedance is presented in Figure \ref{f_diag_regime}(d). This circuit corresponds exactly to Equations \ref{e_impedance_local} to \ref{e_conc_local} using the matrices $\bm{\mathcal{A}}$ and $\bm{\mathcal{C}}$.

\subsection{Extension to other electrode geometries}

In this last section, an example of using the quadrupole formalism to model transient 3D mass transfer is presented. An MEC with an electrode array of 3 elements with different lengths of 5, 2 and 3 mm was built numerically. The electrodes were positioned at $x=0$, 10 and 17 mm and all had the same width, i.e., 0.5 mm. A voltage with a time step profile was simulated by adjusting the value of $k_0$, and the current density at the electrode surface was computed four times.\\

Calculations were performed using $N=50$ frequencies in $y$, $M=21$ elements in the $z$-direction, and at 20 locations in $x$. The Peclet number was set to 800 at an aspect ratio at 12, i.e., in the second regime indicated in Figure \ref{f_impedance}(a). A Damköhler number of 150 was used. The electrode length $1-e^{Lp}$ in Equation \ref{e_convolution_conv_conc} was generalized to an electrode array of Q elements as
\begin{equation}
\mathcal{L}\lbrace f_{array}(x)\rbrace=F(p)\left(\sum_{i=0}^Q (-1)^ie^{-L_ip}\right),
\end{equation}
where $L_i$ is a vector containing the dimension of the electrode array and was $\bm{L}_x=[0, 5, 10, 12, 17, 19]^T$ for the example presented in this paper. The numerical inversion of the Laplace transform was performed using the Den Iseger algorithm based on the fast Fourier transform (FFT) algorithm \cite{DenIseger2006}.\\

\begin{figure}[H]
\centering
\includegraphics[scale=.4]{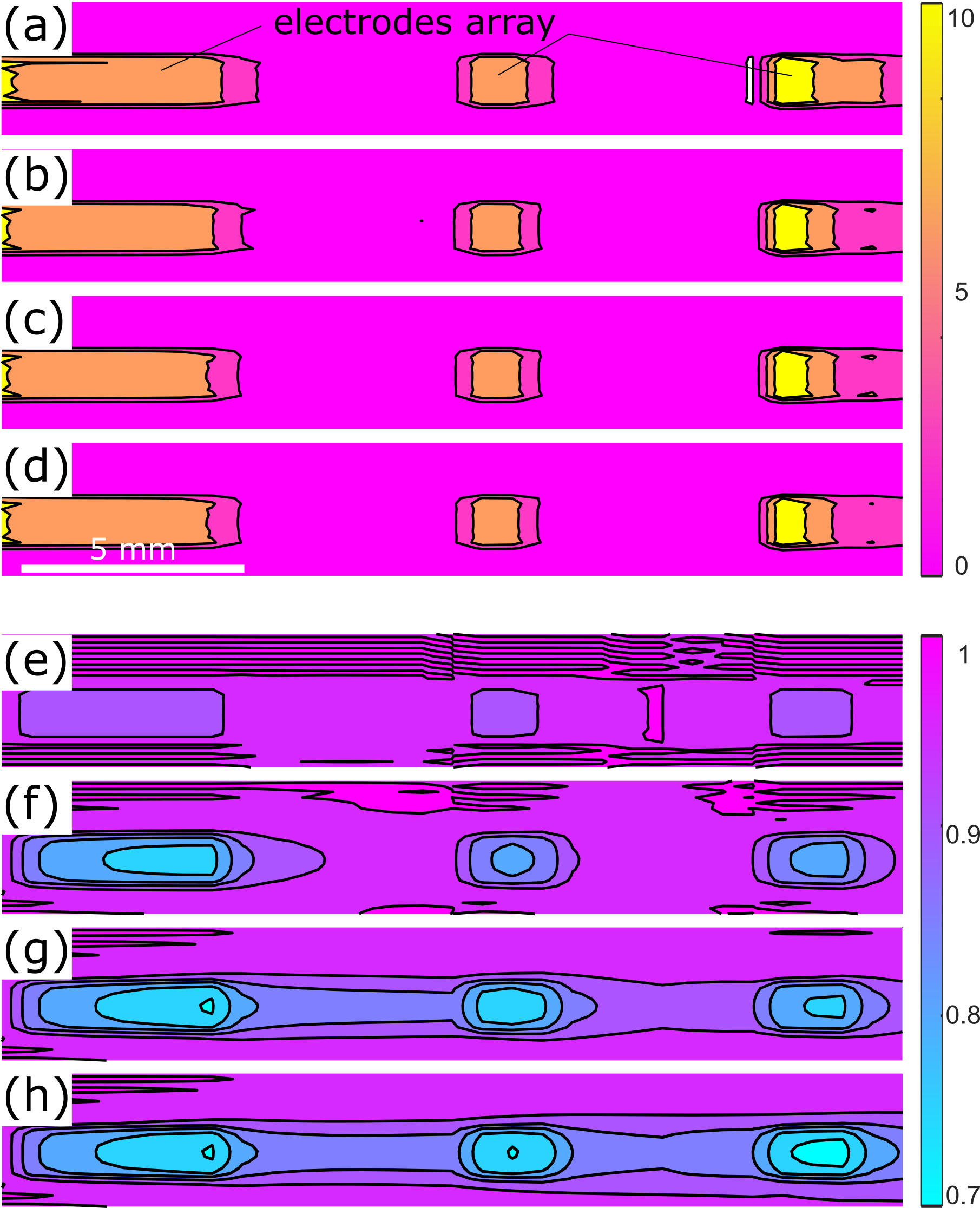}
\caption{Example of calculated fields in an MEC with an electrode array. (a) to (d) Current density fields produced in the MEC at 0, 10, 30 and 60 s, respectively. The color bar unit is µA/cm$^2$. (e)- (f) Dimensionless concentration in the MEC for the same times.}
\label{f_final}
\end{figure}

The simulation results are presented in Figure \ref{f_final}. The subfigures (a) to (d) show the current density in µA computed for a range of time steps. Each area over which the current is different from 0 corresponds roughly to the electrode areas. Semianalytical modeling offers a considerable advantage in requiring computation at only the desired time and location, which reduces the computation time from that of a numerical method in which computation is required at all time steps.\\

In Figure \ref{f_final}(a) to (d), the current density is quite stable over time. A small increase in the current density occurs at times shorter than 60 s, i.e., 0 and 10 s. The current density is only produced in the electrode area. However, in Figure \ref{f_final}(a) to (d), there is some residual current density at the end of the channel. These numerical inaccuracies are mainly due to errors introduced into the numerical inverse Laplace method by using sharp patterns, such as time or spatial steps. Increasing the number of spatial frequencies or $x$-positions may help solve the problem. \\

 Figures \ref{f_final}(e) to (h) show the molar concentration field at $z=0$ for the same time steps. The diffusion and molar concentration consumption in the electrode area can be clearly observed. The effect of advection on the molar concentration can also be seen: the concentration consumed in the first electrodes is transported toward the next electrodes. These results were used to measure the time of flight and estimate the flow rate (see \cite{Kjeang2007}, for example, for a description of this procedure). These calculations yielded the residential time of the chemical species in the channels. The steady states were reached at 60 s. \\

As claimed earlier in the paper, the main advantage of using the proposed semianalytical model is the reduction in the time required to compute 3D transient mass transfer in an MEC: it took only 1 min to compute each time step for the cases presented in Figure \ref{f_final}. However, numerical oscillations appeared due to using the cosine transform, especially in Figure \ref{f_final}(e). Once again, increasing the number of spatial frequencies can solve this problem, but also increases the computation time. However, even with the presence of oscillations in the model solution, the proposed approach remains an effective and fast means of  computing the transient mass transfer and current density in 3D MECs.\\

\section{Conclusions}

A novel formalism for modeling mass transfer in an MEC was derived based on integral transforms. This approach led to a quadrupole formulation of the 3D transient mass transfer equations, which was used to compute the concentration fields and molar flux (or the current density at the electrode interface). The efficiency of using semianalytical solutions was validated against in operando measurements of the concentration field measurements in an MEC. The fast calculation of the concentration fields (less than 1 s) enabled us to use this model in a parameter estimation algorithm to estimate the mass diffusivity and the kinetics reaction rate constant in a few seconds. \\

The main tool developed in this study is a set of impedance models for describing a range of mass transfer regimes from 2D to 3D flow with a parabolic velocity profile. These models meet a strong need for physically based models of the mass transfer impedance for the development of online MEC sensors. The main advantage of these models is the use of a semianalytical solution, which enables the fast and accurate computation of only the fields of interest, i.e., the molar flux or concentration, at a given position and time. \\

This pioneering study using the well-known quadrupole formalism from the heat transfer community to model MEC mass transfer opens up a wide range of possibilities. For example, a periodic solution for mass transfer is induced by an oscillating current or voltage at the electrode interface. This solution reproduces the mass transfer response in well-known electrochemical impedance spectroscopy measurements. Another direction motivated by this study is the development of novel inverse methods to measure the local current density along electrodes. Such studies are currently underway, and the results will be reported in a forthcoming communication.

\section*{Conflicts of interest/Competing interests}
The authors have no conflicts of interest to declare that are relevant to the content of this article.

\section*{Acknowledgement}
The author gratefully acknowledges the French National Research Agency (ANR) for its support through the project I2MPAC, Grant No. ANR-20-CE05-0018-01.

\section*{Appendix}
\subsection*{Calculation of Equation \ref{e_convolution_conv_conc}}
The right side of equation \ref{e_convolution_conv_conc} is in fact the convolution product between the Fourier cosine transform $F(\alpha_n)$ of $f(y)$ and the corresponding cosine transform $(\gamma(h)-\gamma_{lim})$ of the Laplace transform of the concentration $(\Delta\bar{c}-\frac{1}{ps})$. For clarity, this term is simplified as
\begin{equation}
\dot{N}(h)=\int_0^l f(y)\left(\Delta \bar{c}(p,y,z=h,s)-\frac{1}{ps}\right)\cos(\alpha_n y)dy,
\end{equation}
which can be written in vectorial form for a number $N$ of positive harmonics:
\begin{equation}
\dot{N}_n(h)=\sum_{m=0}^{N-1}K_{n-m}(\gamma_m(h)-\gamma_{lim,m}). \label{e_conv_sum}
\end{equation}
Here, keeping only the positive harmonics yields the convolution product. Due to the parity of the cosine transform, it can be shown that $K_{n-m}=K_{-(n-m)}=K_{N-(n-m)}$. Finally, the preceding equation can be written in matrix form as
\begin{equation}
\bm{\dot{N}}(h)=\bm{K}(\bm{\gamma}(h)-\bm{\gamma}_{lim}),
\end{equation}
with the matrix $\bm{K}$ defined as\\
\begin{equation}
\bm{K} =
\begin{bmatrix}
F_0&2F_1&2F_3&\cdots&2F_{N-1}\\
F_1&F_0+F_2&F_{n-m}+F_{n+m}&\cdots&2F_{N-2}\\
F_2&F_{n-m}+F_{n+m}&F_{n-m}+F_{n+m}&\cdots&\vdots\\
\vdots&\vdots&\vdots&\ddots&\vdots\\
F_{N-1}&2F_{N-2}&\cdots&\cdots&2F_0\\
\end{bmatrix}.
\end{equation}
The absence of a factor of  2 in the first column is due the norm of the first harmonic, which is half that of the other harmonics (see Equation \ref{e_inverse_consine_trans}). The sums of the values $F_{n-m}+F_{n+m}$ for all the harmonics represent the convolution product given in Equation \ref{e_conv_sum}.

\subsection*{Calculation of the boundary condition spectra}
The spectra of the functions $\frac{1}{ps}$ and $f(y)$ are given as
\begin{eqnarray}
\gamma_{lim}(\alpha_n)&=&\mathcal{F}_c\left\lbrace \frac{1}{ps}\right\rbrace ,\\
&=&\int_0^{l_c}\frac{\cos(\alpha_ny)}{ps}dy,\\
&=& \frac{\delta_c}{ps},
\end{eqnarray}
where $\delta_c=l_c$ for $n=0$ and 0 for all the other harmonics ($n\geq 1$), and
\begin{eqnarray}
F(\alpha_n)&=&\mathcal{F}_c\left\lbrace f(y)\right\rbrace ,\\
&=&\int_0^{l_c}\Theta(y-l_1)\Theta(l_1+e-y)\cos(\alpha_ny)dy,\\
&=& \frac{\sin(\alpha_n(l_1+e))-\sin(\alpha_nl_1)}{\alpha_n},
\end{eqnarray}
with $F(\alpha_0)=e$. Note that this calculation was performed for the case of a rectangular electrode. The spectrum needs to be recomputed for a different electrode geometry.\\
The inverse cosine transform can then be used, which is given for a finite number of harmonics $N$ by
\begin{equation}
\Delta \bar{c} = \frac{1}{l_c}\gamma_0(h)+\frac{2}{l_c}\sum_{n=1}^N\gamma_n(h)\cos(\alpha_ny) .\label{e_inverse_consine_trans}
\end{equation}

\subsection*{Calculation of the coefficients $\bm{\mathcal{A}}(z)$, $\bm{\mathcal{B}}(z)$ and $\bm{\mathcal{C}}(z)$}
The coefficients used to compute the concentration profile at a location $z=qdh$ are extracted from the following matrix:
\begin{equation}
\begin{bmatrix}
\bm{\mathcal{A}}(z)&\bm{\mathcal{B}}(z)\\
\bm{\mathcal{C}}(z)&\bm{\mathcal{A}}(z)
\end{bmatrix}
=
\prod_{i=1}^q
\begin{bmatrix}
\bm{A}_i&\bm{B}_i\\
\bm{C}_i&\bm{A}_i
\end{bmatrix},
\end{equation}
where the coefficients $\bm{A}_i$, $\bm{B}_i$ and $\bm{C}_i$ are given by Equations \ref{e_coef_Ai} to \ref{e_coef_Ci}.
\clearpage

\bibliographystyle{unsrt}
\bibliography{references}  






\end{document}